# An Approach for Message Hiding using Substitution Techniques and Audio Hiding in Steganography

Dr D Mukhopadhyay, *Fellow*
A Mukherjee, *Non-member*
S Ghosh, *Non-member*
S Biswas, *Non-member*
P Chakraborty, *Non-member*

*A crypto system can be used to encrypt messages sent between two communicating parties so that an eavesdropper who overhears the encrypted messages will not be able to decode them. The paper mainly concentrates on the method in which the substitution technique of steganography can been used to hide data in a 24-bit bitmap file. Popular audio hiding techniques based on methods of steganography is also discussed here.*

## INTRODUCTION

Information hiding started its development as a separate sub discipline of information security with steganography[1] and watermarking as its main branches. It borrowed ideas from the more established sister disciplines, namely, cryptography[2]. A crypto system can be used to encrypt messages sent between two communicating parties so that an eavesdropper who overhears the encrypted messages will not be able to decode them.

The paper mainly concentrates on the method in which the substitution technique of steganography can been used to hide data in a 24-bit bitmap file and try to focus on the popular audio hiding techniques based on methods of steganography.

## THE ALGORITHM

**Process of Encryption**

The following section tries to develop a sample code to show the effect of hiding a secret message in an image large enough to cover the message.

An abridged version of the algorithm used is

*Cover Media* : A 24-bit BMP file ( File name: C).

*Secret Message* : Can be text, audio, video or any other file format (File name: M).

*Algorithm* :

1. Start.
2. The RGB information is read from the cover file C into bytes r, g and b.
3. Two bits from the LSB of r is replaced by the first two bits of the first character in the secret message file M.
4. Three bits from the LSB of g is replaced by the next three bits of the first character in M.
5. Three bits from the LSB of b is replaced by the next three bits of the first character in M. In this way one byte of M gets covered by the r, g, b bytes of C.
6. The next r, g and b bytes are chosen from the cover file C after leaving a fixed amount of space which equals (cover file size)/size of (r+g+b)*(message file size).
7. Steps 3 to 6 are repeated throughout the file C until the entire message gets embedded.

The information about the message file size and the extension of the message file to be formed after decryption at the receiver end also remains hidden in the encrypted file. The encrypted file also bears information whether it is a genuine encryption or not.

**Process of Decryption**

The decryption is done in just a process similar to that of encryption.

*Encrypted Message or the Stego Media* : A 24-bit bitmap file (File name: E).

*Algorithm*:

1. Start.
2. At first the file E is tested to know whether it is a genuine encryption or not.
3. Next while reading the header of E, the size of the message file is obtained from the reserved bits and the pixel spacing is calculated using the same formula as in the case of encryption.
4. The extension information of the message file is read from the header of E and accordingly an empty file is built with the same extension that serves as the destination.
5. Next the r, g, and b bytes are read from E at equal pixel spacing and the message bits are extracted and stored sequentially in the destination file created above.
6. Step 5 is continued until the end of file of E is encountered.

Dr D Mukhopadhyay is with Cellular Automata Research Laboratory (CARL) , Techno India, Kolkata 700 091 ; A Mukherjee is with Indian Institute of Technology, Kharagpur 721 302;  S Ghosh is with Tata Consultancy Services; S Biswas is with Institute of Technology and Marine Engineering while  P Chakraborty is with Infosys Technologies.

## Mathematics of Encryption

Let us consider the case of encrypting the character 'a'.

Now ASCII('a') = 97

The binary form is :

0 1 1 0 0 0 0 1

Let the r, g, b values be as shown :

```
       r                g                b
1 0 0 1 0 0 1 1   1 1 0 1 0 1 0 1   1 0 1 1 0 0 1 1
```

Let ch_r=(ch & 192)>>6 = (01000000)>>6 = (00000001).
Let ch_g=(ch & 56)>>3 = (00100000)>>3 = (00000100).
Let ch_b=(ch & 7) = (00000001).

r = r & 252 = (10010000).
r = r | ch_r = (10010001).

g = g & 248 = (11010000).
g = g | ch_g = (11010100).

b = b & 248 = (10110000).
b = b | ch_b = (10110001).

```
       r                g                b
1 0 0 1 0 0 0 1   1 0 0 1 0 1 0 0   1 0 0 1 0 0 0 1
```

## Mathematics of Decryption

Let ch = '\0' ⟶    0 0 0 0 0 0 0 0

r = r & 3 = (00000001).
g = g & 7 = (00000100).
b = b & 7 = (00000001).
ch = ch | (r << 6) = (00000000) | (01000000) = (01000000).
ch = ch | (g << 3) = (01000000) | (00100000) = (01100000).
ch = ch | b = (01100000) | (00000001) = (01100001).

ch ⟶    0 1 1 0 0 0 0 1

Hence one have ch = 'a', since ASCII(ch) = 97

A program shown in appendix has been developed in 'C' to implement this algorithm.

## STEGANOGRAPHY IN AUDIO

As an integral backbone of voice communication, the internet provides various audio applications like voice query, voice activated websites, etc. On the internet, the windows audio visual (WAV), audio interchange file format (AIFF) and motion picture experts group layer III (MP3) files support a data rate varying from 8 kbps to 44.1 kbps. Audio traffic on the internet system is increasing rapidly, that's why it is obvious to choose an audio file as a cover media.

Audio hiding techniques rely on the weakness of the human auditory system (HAS). Small variation in the amplitude level, phase, spectral magnitude and some other parameters can be done without substantially degrading the signal. LSB insertion can be easily done to generate high embedding capacity in audio signals. The noise introduced due to embedding can be adjusted using adaptive data attenuation.[3]

### Popular Techniques of Audio Encryption

There are various techniques of audio encryption with phase coding and echo hiding as its main sub categories.

*Phase Coding*

Small modification in the phase of the signal is done in such a way that the relative phase among segment is preserved. As for example, the cover signal has already been transmitted to the receiver and now the same signal with some delay in the phase is being transmitted as given in Figure 1. Here the receiver has to follow the two incoming signals to recover the original message that is being embedded into the phase of the cover signal.

*Echo Hiding*

Here an echo is introduced in a signal by manipulating the parameters (initial amplitude, decay and offset) such that the echo is not audible. For a discrete signal f(t), an echo f(t-dt), with some delay can be introduced to produce the stego signal s(t)=f(t) + f(t-dt). Figure 2 represent the same.

By two procedures it can be explained how the echo is embedded and how it is extracted from cover signal so as to recover the data. Firstly, it can be explained by means of convolution theory. As for example, a non return to zero (NRZ) cover signal has been transmitted to the receiver and now the cover signal with some delay (echo) is again transmitted. Now the convolution theory speaks that, when the two signals are overlapped then there will be some state where the two nullify each other. The duration of the nullification will express the delay in which transmitter has stored the hidden message. Here f(t) is the cover signal and dt is delay as well as the hidden message, as given in Figure 3.

Another way can be used to explain the technique which is shown in Figure 4. In this case one performs a modulo two addition (XOR) of the signal f(t) and shifted signal f(t-dt). The ultimate signal is then transmitted. The receiver has already been informed that a XOR operation is performed. The receiver has to pass the signal through a toggle-flip flop (positive edge-triggered) to recover the original signal dt.

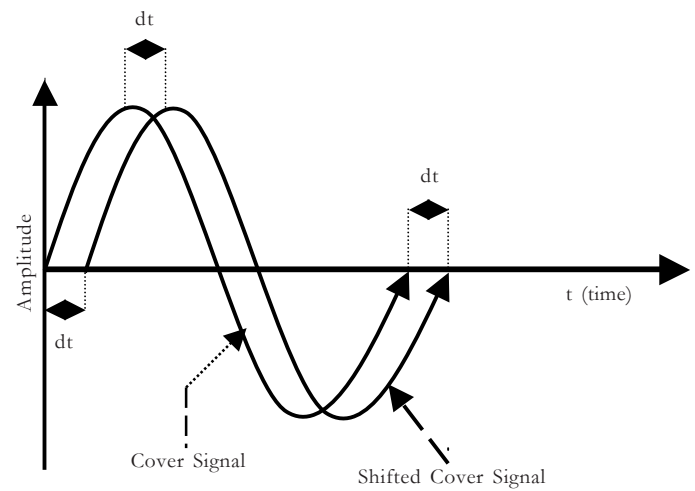

Figure 1 Phase coding

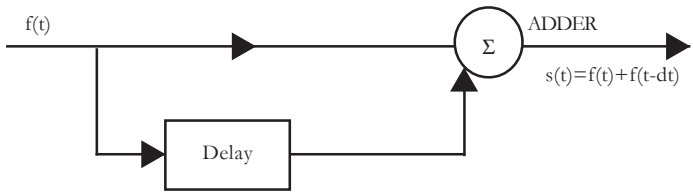

**Figure 2 Echo hiding**

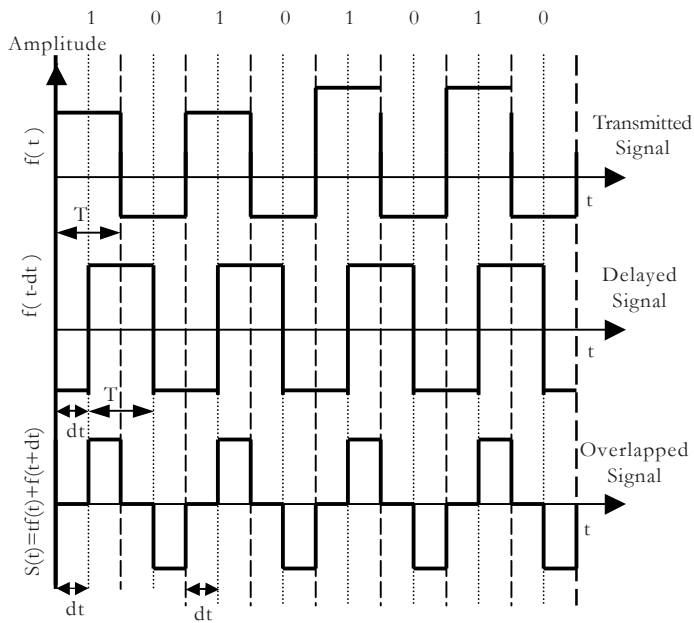

**Figure 3 Echo hiding through convolution**

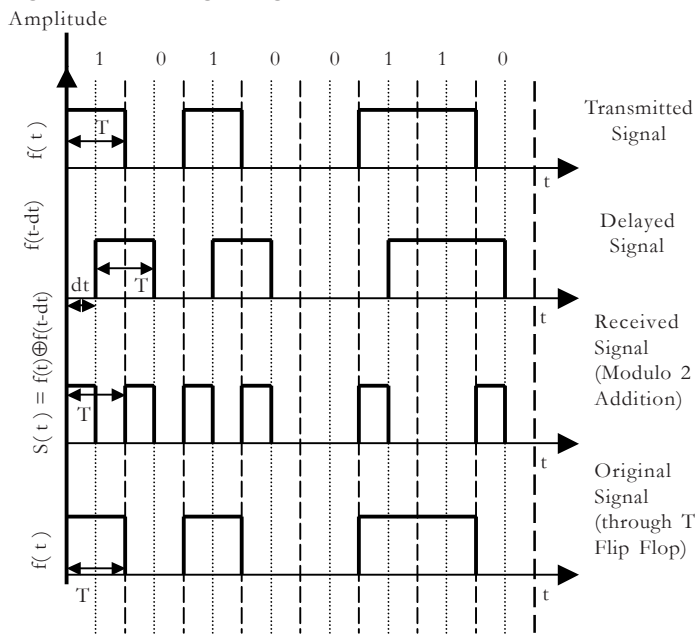

**Figure 4 Echo hiding through modulo (2) addition**

## CONCLUSION

A brief description of different steganographic techniques used for information hiding has been given in this article. The choice of the technique depends upon the application scenario and takes into consideration factors like robustness to attacks, the data-capacity of the secret message[4] and the perceivability of the stego media. Cryptography may be used as an additional security layer over information hiding but should not give any clue for message detection . Proper use of information hiding techniques for military as well as domestic applications help to provide safe communication during war and peace .

## REFFERENCES

## APPENDIX

**Program to Encrypt and Decrypt any File Into a BMP File**

```
#include<stdio.h>
#include<conio.h>
#include<ctype.h>
#include<string.h>
#include<stdlib.h>

struct bitmap
{ char identity[2];
long f_size; long reserved;
long offset; long head_size;
long wid; long height;
int plane; int bpp; };

struct rgb
{ char r,g,b; };

long int find_char(char[ ]);
void extract_extn(char*,char*);
void encrypt_img(char[ ],char[ ]);
void decrypt_img(char[ ]);

void main()
{ char ch='y', msg_file[50], cover_file[50];
int opt,enc_opt,dec_opt;
while((toupper(ch))=='Y')
{ clrscr(); printf("\n\n\n\t 1:Encryption...\n\t\t 2:Decryption...\n\t\t 3:Exit.");
printf("\n\t\t\t Enter your choice: "); scanf("%d",&opt);
if(opt==1)
{ clrscr(); printf("\n\n\n\t Enter the message file name: ");
scanf("%s",msg_file);
strcpy(cover_file,"c:\\sun\\seminer\\cvr.bmp");
encrypt_img(msg_file,cover_file); }
else if(opt==2)
{ clrscr(); printf("\n\n\n\t Enter the encrypted file name: ");
scanf("%s",msg_file);
decrypt_img(msg_file); }
else if(opt==3) exit(1); }
printf("\n\n\t\t\t Do you want to continue(y/n): ");
ch=getche(); }}
long find_char(char file_name[])
{ char ch; long count=0;
```

```c
FILE *fp; fp=fopen(file_name,"rb");
while(fread(&ch,sizeof(ch),1,fp)==1)
count++; fclose(fp);

return count; }
void extract_extn(char *str,char *extn)
{ int i=0,k;
for(i=0;i<strlen(str);i++)
{ if(str[i]!='.') continue;
else { i=i+1;
for(k=i;k<strlen(str);k++)
{ extn[k-i]=str[k]; }
extn[k-i]='\0'; }}}
void encrypt_img(char msg_file[ ],char cover_file[ ])
{ char ch, ch_r, ch_g, ch_b, extn[5], enc_file[ ]="c:\\sun\\seminer\\enc.bmp";
int i,x,y; long file_size,pixel_spacing,gap=0;
struct bitmap cvr; struct rgb cvr_color;
FILE *fp_msg,*fp_cvr,*fp_enc;
file_size=find_char(msg_file);
extract_extn(msg_file,extn);
fp_msg=fopen(msg_file,"rb");
fp_cvr=fopen(cover_file,"rb");
fp_enc=fopen(enc_file,"wb");
fread(&cvr,sizeof(cvr),1,fp_cvr);
pixel_spacing=(cvr.f_size-cvr.offset)/(sizeof(struct rgb)*file_size);
cvr.reserved=file_size;
fwrite(&cvr,sizeof(cvr),1,fp_enc);
rewind(fp_enc);
fseek(fp_enc,cvr.head_size,SEEK_CUR);
ch='1'; fwrite(&ch,sizeof(ch),1,fp_enc);
for(i=0;i<strlen(extn);i++)
fwrite(&extn[i],sizeof(extn[i]),1,fp_enc);
rewind(fp_cvr); rewind(fp_enc);
fseek(fp_cvr,cvr.offset,SEEK_CUR);
fseek(fp_enc,cvr.offset,SEEK_CUR);
for(y=cvr.height-1;y>=0;y--)
{for(x=0;x<cvr.wid;x++)
{fread(&cvr_color,sizeof(cvr_color),1,fp_cvr);
if(gap==pixel_spacing)
{ if( (fread(&ch,sizeof(ch),1,fp_msg) )==1)
{ ch_r=(ch&192)>>6; ch_g=(ch&56)>>3; ch_b=ch&7;
cvr_color.r=cvr_color.r&252;
cvr_color.r=cvr_color.r|ch_r;
cvr_color.g=cvr_color.g&248;
cvr_color.g=cvr_color.g|ch_g;
cvr_color.b=cvr_color.b&248;
cvr_color.b=cvr_color.b|ch_b;
} gap=0; }
fwrite(&cvr_color,sizeof(cvr_color),1,fp_enc);
gap++; }}
fclose(fp_msg); fclose(fp_cvr); fclose(fp_enc);
printf("\n\n\n\t\t Encryption is done successfully...");
printf("\n\n\n\t\tPath of the encrypted file: %s",enc_file); }
void decrypt_img(char msg_file[ ])
{ char ch, extn[5], dec_file[50]="c:\\sun\\seminer\\org.";
int i,x,y; long pixel_spacing,count=0,file_size;
FILE *fp_msg,*fp_dec; struct bitmap msg;
struct rgb msg_color; fp_msg=fopen(msg_file,"rb");
fread(&msg,sizeof(msg),1,fp_msg); rewind(fp_msg);
fseek(fp_msg,msg.head_size,SEEK_CUR);
fread(&ch,sizeof(ch),1,fp_msg);
if(ch!='1')
{ printf("\n\n\t\t\tUnrecognised Format...Can't be decrypted...");
printf("\n\n\t\t\t Press any key to exit...");
getche(); exit(0); }
for(i=0;i<3;i++)
{ fread(&extn[i],sizeof(extn[i]),1,fp_msg); }
extn[i]='\0'; strcat(dec_file,extn);
fp_dec=fopen(dec_file,"wb");
file_size=msg.reserved;
pixel_spacing=(msg.f_size-msg.offset)/(sizeof(struct rgb)*file_size);
rewind(fp_msg); fseek(fp_msg,msg.offset,SEEK_CUR);
fseek(fp_msg,pixel_spacing*sizeof(struct rgb),SEEK_CUR);
count=0; while(1)
{ fread(&msg_color,sizeof(msg_color),1,fp_msg);
fseek(fp_msg,(pixel_spacing-1)*sizeof(struct rgb),SEEK_CUR);
ch='\0'; msg_color.r=msg_color.r&3;
msg_color.g=msg_color.g&7;
msg_color.b=msg_color.b&7;
ch=ch|(msg_color.r<<6);
ch=ch|(msg_color.g<<3);
ch=ch|msg_color.b;
fwrite(&ch,sizeof(ch),1,fp_dec);
count++;
if(count==file_size)
break; }
fclose(fp_msg); fclose(fp_dec);
printf("\n\n\n\n\t\t\t Decryption is done successfully...");
printf("\n\n\t\tPath of the decrypted file: %s",dec_file); }
}
```